# Design and Performance Analysis of a Class of Generalized Predictive Controllers

Feilong Zhang
State Key Laboratory of Robotics, Shenyang Institute of Automation, Chinese Academy of Sciences,
Shenyang 110016, China

*Abstract*—The design and structure of generalized predictive control (GPC) are not simple and intuitive. The performance analysis does not deeply analyze how the controller parameters affect the system characteristics and the relationship between the tracking error caused by the noise and the selected controller parameters. This paper proposes a generalized predictive control, and its design is simple and intuitive for unnecessary solving the Diophantine equation. Then the relationship between desired output, disturbance, and system output is analyzed by the characteristic equation and steady-state analysis. Based on this, the study presents research findings on the steady state of the system and verifies them through simulations. Furthermore, this paper introduces GPC with disturbance compensation and incremental generalized minimum variance control (IGMVC) with disturbance compensation. The conditions for the elimination of disturbance are presented in theory and simulation for the first time.

## I. INTRODUCTION

In 1973, Åström and others introduced the minimum variance self-tuning regulator. Due to its simplicity, ease of understanding and implementation, this algorithm garnered widespread attention. The key to implementing minimum variance control lies in predicting the output of the model. Typically, controlled processes exhibit a time delay, denoted as '$d$', and the control action at the current instant affects the system's output only after a lag of '$d$' sampling times [1]-[4]. To minimize the variance of the output tracking error, it is necessary to make '$d$' step-ahead predictions of the output variable. Subsequently, the required control action can be designed based on these predictions and the objective function. In the case of non-minimum phase systems, minimum variance control leads to issues of unbounded control inputs. Clarke introduced a weighting term on the control action to constrain its magnitude or rate of change, which gave rise to the concept of generalized minimum variance self-tuning control [5]. Building upon this, Clarke introduced the concept of generalized predictive control (GPC) by incorporating rolling optimization strategies similar to DMC and MAC [6]-[10]. This algorithm has found applications in various fields, including process industries, atomic force microscopy systems, motor servo systems, high-speed trains, medical systems, and more [11]-[16]. Moreover, it has been combined with neural networks or fuzzy control for industrial process control applications [17]-[19].

Goodwin introduced the principle of certainty equivalence in [20]: "In concept, the adaptive control system is simple. A natural way is to combine a specific parameter estimation method with any control law and treat the estimated values as if they were the true parameters for the design purposes [20]." Numerous scholars have designed controllers and analyzed system performance under this principle. They initially design a generalized predictive controller using actual model values and then perform performance analysis on the system. Scholars like Wang and Chai [21]-[26], following the [20] approach, designed generalized predictive control algorithms and analyzed the stability or performance of the system, assuming that the disturbance term is white noise. The conclusion drawn from this was that the upper bound of the system tracking error is the mean of the noise term's function. However, these results did not provide a clear description of the relationship between tracking errors caused by noise terms and the selected controller parameters. Xi et al. analyzed the closed-loop properties of generalized predictive control and provided several stability results [27]-[31]. However, much of the information was not intuitively presented. One reason for this is that the design from minimum variance control to generalized predictive control involves solving Diophantine equations, making it difficult to intuitively show how model parameters and controller parameters affect system characteristics. The author has not come across any scholars who have analyzed the steady-state characteristics of ramp input and acceleration input according to[32], [33], which is essential for control algorithm research. Cordero et al. improved generalized predictive control by introducing a second integrator to increase the system type and improve tracking performance for ramp signals. However, serially integrating two integrators may lead to integrator saturation, and the choice of system characteristics obtained through tuning will be very narrow and often not the desired system performance [34], [35].

This paper builds upon previous literature on controlled autoregressive integrated moving average (CARIMA) models and the objective function in order to redesign the generalized predictive control. (1) Instead of deriving the prediction model through the traditional Diophantine equation, this work directly deduces it from CARIMA. This approach offers several advantages: it simplifies controller design and provides a more explicit relationship between each element in the controller's coefficient matrix and the model. The system characteristic equation also offers a more intuitive depiction of the relationship between reference input, disturbances, and system output. (2) Furthermore, this research introduces the incremental generalized minimum variance control (IGMVC) in Section 4. This variant of control provides a clearer understanding of the controller's effects. (3) The current literature on GPC does not explicitly address the steady-state analysis of reference inputs, including slope reference input $k$, acceleration reference input $k^2$, and higher-order reference inputs $k^n$ ($n=3, \cdots$). This paper bridges that gap by conducting a stability analysis of practical reference inputs and validating the findings through simulations. Additionally, this work provides theoretical evidence that the simplified controller discussed in Section 6 can achieve zero steady-state error when tracking reference input signals $k^n$ ($n=1, 2, \cdots$), under the condition of a controller parameter matrix $\boldsymbol{\lambda}=\boldsymbol{0}$ and the stability of the system. Moreover, simulations provide insights into the scenarios where the diagonal elements in $\boldsymbol{\lambda}$ can be negative. This elucidation is done within the context of the corresponding system characteristics.

(4) Section 5 of this paper introduces disturbance compensation for generalized predictive control. Theoretical conditions and simulation-based evidence are provided to illustrate the effective elimination of the impact of known disturbances on the system output.

(5) Finally, the paper simplifies the form of the proposed generalized predictive controller, reducing the dimensionality of matrix inversions in predictive control. This reduction brings about computational benefits, as computational workload decreases with diminishing matrix dimensionality.

## II. SYSTEM DESCRIPTION

Consistent with references [9], [21], [25], consider the controlled object described by the CARIMA model as follows

$$A(z^{-1})y(k) = B(z^{-1})u(k-1) + \chi(k) \quad (1)$$

where, $A(z^{-1})$, $B(z^{-1})$, and $C(z^{-1})$ are polynomials of the backward shift operator $z^{-1}$. $\chi(k)$ denotes disturbance, specifically referred to in references [9] and [21] as follows

$$\chi(k) = C(z^{-1})\zeta(k)/\Delta \quad (2)$$

where $C(z^{-1})=1+c_1 z^{-1}+\cdots+c_{nc} z^{-n_c}$.

For details, please refer to reference [25], where $u(k)$ and $y(k)$ represent the inputs and outputs of the system. $\Delta=1-z^{-1}$ represents a difference operator.

If the estimated parameters converge to a certain value, then for a general ARMAX model, even if the noise parameters are not explicitly estimated, the self-tuning minimum variance controller will still be optimal [20]. Following the design approach in [25], assuming that $c_i$ is unknown, the controller designed without processing the noise term as $\chi(k)$ results in overlapping controlled system outputs when compared to the controller designed with the true values of $c_i$. Therefore, in the following sections, the disturbance term is simplified as $\chi(k)$.

In contrast to references [9], [21] and [25], this paper does not impose any restrictions on disturbances. The primary objective is to explore and analyze the broader range of impacts that disturbances have on the system output.

Differencing both sides of model (1), we can obtain

$$A(z^{-1})\Delta y(k) = B(z^{-1})\Delta u(k-1) + \Delta \chi(k) \quad (3)$$

## III. MODEL PREDICTION

We can rewrite (3) into state space form
$$\begin{aligned}\Delta \boldsymbol{x}(k+1) &= \boldsymbol{A}\Delta \boldsymbol{x}(k) + \boldsymbol{B}\Delta u(k) + \boldsymbol{T}\Delta \chi(k) \\ \Delta y(k+1) &= \boldsymbol{C}\Delta \boldsymbol{x}(k+1)\end{aligned} \quad (4)$$

where
$\Delta \boldsymbol{x}(k) = [\Delta y(k),\cdots,\Delta y(k-n_a+1), \Delta u(k-1),\cdots,\Delta u(k-n_b-1)]^T$
$\boldsymbol{B}^T = [b_0, 0,\cdots,0,1,0,\cdots,0]_{1\times(na+nb+1)}$, $\boldsymbol{C} = \boldsymbol{T}^T = [1,0,\cdots,0]$

$$\boldsymbol{A} = \begin{bmatrix} -a_1 & -a_2 & \cdots & -a_{na-1} & -a_{na} & b_1 & \cdots & b_{nb}(k) & 0 \\ 1 & 0 & \cdots & 0 & 0 & 0 & \cdots & 0 & 0 \\ 0 & 1 & \cdots & 0 & 0 & 0 & \ddots & 0 & 0 \\ \vdots & \vdots & \ddots & \vdots & \vdots & \vdots & \ddots & \vdots & \vdots \\ 0 & 0 & \cdots & 1 & 0 & 0 & \cdots & 0 & 0 \\ 0 & 0 & \cdots & 0 & 0 & 0 & \cdots & 0 & 0 \\ 0 & 0 & \cdots & 0 & 0 & 1 & \ddots & 0 & 0 \\ \vdots & \vdots & \ddots & \vdots & \vdots & \vdots & \ddots & \vdots & \vdots \\ 0 & 0 & \cdots & 0 & 0 & 0 & \cdots & 1 & 0 \end{bmatrix}$$

We perform N-step prediction on model (4) to obtain

$$\begin{aligned}\Delta \boldsymbol{x}(k+1) &= \boldsymbol{A}\Delta \boldsymbol{x}(k) + \boldsymbol{B}\Delta u(k) + \boldsymbol{T}\Delta \chi(k+1) \\ \Delta \boldsymbol{x}(k+2) &= \boldsymbol{A}\Delta \boldsymbol{x}(k+1) + \boldsymbol{B}\Delta u(k+1) + \boldsymbol{T}\Delta \chi(k+2) \\ &= \boldsymbol{A}\boldsymbol{A}\Delta \boldsymbol{x}(k) + \boldsymbol{A}\boldsymbol{B}\Delta u(k) \\ &\quad + \boldsymbol{B}\Delta u(k+1) + \boldsymbol{A}\boldsymbol{T}\Delta \chi(k+1) + \boldsymbol{T}\Delta \chi(k+2) \\ &\vdots \\ \Delta \boldsymbol{x}(k+N) &= \boldsymbol{A}\Delta \boldsymbol{x}(k+N-1) + \boldsymbol{B}\Delta u(k+N-1) + \boldsymbol{T}\Delta \chi(k+N) \\ &= \boldsymbol{A}^N \Delta \boldsymbol{x}(k) + \boldsymbol{A}^{N-1}\boldsymbol{B}\Delta u(k) + \boldsymbol{A}^{N-2}\boldsymbol{B}\Delta u(k+1) + \cdots \\ &\quad + \boldsymbol{B}\Delta u(k+N-1) + \boldsymbol{A}^{N-1}\boldsymbol{T}\Delta \chi(k+1) \\ &\quad + \boldsymbol{A}^{N-2}\boldsymbol{T}\Delta \chi(k+2) + \cdots + \boldsymbol{T}\Delta \chi(k+N)\end{aligned} \quad (5)$$

where $N$ represents the prediction time horizon, $\Delta y(k+i)$ and $\Delta u(k+i)$ represent the increments of the system output and input at time $k+i$ ($i=1,2,\cdots,N$), respectively. We define $\boldsymbol{Y}_N(k)$, $\Delta \boldsymbol{Y}_N(k+1)$, $\Delta \boldsymbol{U}_N(k)$, $\boldsymbol{\Psi}$, $\tilde{\boldsymbol{\Psi}}$, $\boldsymbol{\Phi}$, $\tilde{\boldsymbol{\Phi}}$, $\boldsymbol{\Phi}_w$ 和 $\tilde{\boldsymbol{\Phi}}_w$ as follows:

$$\boldsymbol{\Phi} = \begin{bmatrix} CB & 0 & \cdots & 0 \\ CAB & CB & \cdots & 0 \\ \vdots & \vdots & \ddots & \vdots \\ CA^{N-1}B & CA^{N-2}B & \cdots & CB \end{bmatrix},$$

$$\tilde{\boldsymbol{\Phi}} = \boldsymbol{\Lambda}_N \boldsymbol{\Phi} = \begin{bmatrix} CB & 0 & \cdots & 0 \\ CAB+CB & CB & \cdots & 0 \\ \vdots & \vdots & \ddots & \vdots \\ \sum_{j=0}^{N-1} CA^j B & \sum_{j=0}^{N-2} CA^j B & \cdots & CB \end{bmatrix},$$

$$\boldsymbol{\Psi} = \begin{bmatrix} CA \\ CAA \\ \vdots \\ CA^N \end{bmatrix}, \quad \tilde{\boldsymbol{\Psi}} = \boldsymbol{\Lambda}_N \tilde{\boldsymbol{\Psi}} = \begin{bmatrix} CA \\ CAA+CA \\ \vdots \\ \sum_{i=1}^{N} CA^i \end{bmatrix},$$

$$\boldsymbol{\Phi}_w = \begin{bmatrix} CT & 0 & \cdots & 0 \\ CAT & CT & \cdots & 0 \\ \vdots & \vdots & \ddots & \vdots \\ CA^{N-1}T & CA^{N-2}T & \cdots & CT \end{bmatrix}, \quad \boldsymbol{E} = \begin{bmatrix} 1 \\ \vdots \\ 1 \end{bmatrix}_{N \times 1},$$

$$\tilde{\boldsymbol{\Phi}}_w = \boldsymbol{\Lambda}_N \boldsymbol{\Phi}_w =$$

$$\begin{bmatrix} CT & 0 & \cdots & 0 \\ CAT+CT & CT & \cdots & 0 \\ \vdots & \vdots & \ddots & \vdots \\ \sum_{i=0}^{N-1} CA^i T & \sum_{i=0}^{N-2} CA^i T & \cdots & CT \end{bmatrix}, \quad \Delta\boldsymbol{\chi}(k+1) = \begin{bmatrix} \Delta\chi(k+1) \\ \vdots \\ \Delta\chi(k+N) \end{bmatrix},$$

$$\boldsymbol{Y}_N(k+1) = \begin{bmatrix} y(k+1) \\ \vdots \\ y(k+N) \end{bmatrix}_{N \times 1}, \quad \Delta\boldsymbol{U}_N(k) = \begin{bmatrix} \Delta u(k) \\ \vdots \\ \Delta u(k+N-1) \end{bmatrix}_{N \times 1},$$

$$\boldsymbol{\Lambda}_N = \begin{bmatrix} 1 & & \\ \vdots & \ddots & \\ 1 & \cdots & 1 \end{bmatrix}_{N \times N}, \quad \Delta\boldsymbol{Y}_N(k+1) = \boldsymbol{Y}_N(k+1) - \boldsymbol{Y}_N(k).$$

We can express (5) as (6)

$$\Delta\boldsymbol{Y}_N(k+1) = \boldsymbol{\Psi}\Delta\boldsymbol{x}(k) + \boldsymbol{\Phi}\Delta\boldsymbol{U}_N(k) + \boldsymbol{\Phi}_w \Delta\boldsymbol{\chi}(k+1) \quad (6)$$

By left-multiplying both sides of (6) with $\boldsymbol{\Lambda}_N$, we obtain

$$\boldsymbol{Y}_N(k+1) = \boldsymbol{E}y(k) + \boldsymbol{\Lambda}_N \boldsymbol{\Psi} \Delta\boldsymbol{x}(k) + \boldsymbol{\Lambda}_N \boldsymbol{\Phi} \Delta\boldsymbol{U}_N(k)$$
$$+ \boldsymbol{\Lambda}_N \boldsymbol{\Phi}_w \Delta\boldsymbol{\chi}(k+1) \quad (7)$$
$$= \boldsymbol{E}y(k) + \tilde{\boldsymbol{\Psi}}\Delta\boldsymbol{x}(k) + \tilde{\boldsymbol{\Phi}}\Delta\boldsymbol{U}_N(k) + \tilde{\boldsymbol{\Phi}}_w \Delta\boldsymbol{\chi}(k+1)$$

## IV. GENERALIZED PREDICTIVE CONTROLLERS DESIGN AND PERFORMANCE ANALYSIS

### A. Predictive Controllers Design and Performance Analysis

Design the objective function as:

$$J = \left[\boldsymbol{Y}_N^*(k+1) - \boldsymbol{Y}_N(k+1)\right]^T \boldsymbol{Q} \left[\boldsymbol{Y}_N^*(k+1) - \boldsymbol{Y}_N(k+1)\right]$$
$$+ \Delta\boldsymbol{U}_N^T(k) \boldsymbol{\lambda} \Delta\boldsymbol{U}_N(k) \quad (8)$$

where $\tilde{\boldsymbol{Y}}_N^*(k+1) = \left[y^*(k+1), \cdots, y^*(k+N)\right]^T$ is the desired output vector of the system in future time. $y^*(k+i)$ is the expected output of the system at time $k+i$ ($i=1,2,\cdots,N$). $\boldsymbol{\lambda}=\text{diag}(\lambda_1,\cdots,\lambda_N)$ and $\boldsymbol{Q}=\text{diag}(q_1,\cdots,q_N)$ are diagonal weight matrices. For the convenience of stability proof or system performance analysis, references [9], [21], [25] specify the parameters $q_i=1$ ($i=1,\cdots,N$), $\lambda_i=\lambda$ ($i=1,\cdots,N$). This paper does not impose any restrictions on the range of $\lambda$. Additionally, the analytical approach utilized in this study provides a more intuitive understanding of the impact of $\lambda$ and $\boldsymbol{Q}$ on system characteristics.

By incorporating (7) into (8) and minimizing the objective function (8), the result obtained is

$$\Delta\boldsymbol{U}_N(k) = [\tilde{\boldsymbol{\Phi}}^T \boldsymbol{Q} \tilde{\boldsymbol{\Phi}} + \boldsymbol{\lambda}]^{-1} \tilde{\boldsymbol{\Phi}}^T \boldsymbol{Q}[\boldsymbol{Y}_N^*(k+1)$$
$$- \boldsymbol{E}y(k) - \tilde{\boldsymbol{\Psi}} \Delta\boldsymbol{x}(k) - \tilde{\boldsymbol{\Phi}}_w \Delta\boldsymbol{\chi}(k+1)] \quad (9)$$

Since [25] assumes the disturbance is unknown, we set $\Delta\boldsymbol{\chi}(k+1)=\boldsymbol{0}$, which allows us to rewrite equation (9) as:

$$\Delta\boldsymbol{U}_N(k) = [\tilde{\boldsymbol{\Phi}}^T \boldsymbol{Q} \tilde{\boldsymbol{\Phi}} + \boldsymbol{\lambda}]^{-1} \tilde{\boldsymbol{\Phi}}^T \boldsymbol{Q}[\boldsymbol{Y}_N^*(k+1)$$
$$- \boldsymbol{E}y(k) - \tilde{\boldsymbol{\Psi}}(k) \Delta\boldsymbol{x}(k)] \quad (10)$$

We take the first component of it, which is the current control output, and send it to the controlled object.

$$u(k) = u(k-1) + \boldsymbol{g}^T \Delta\boldsymbol{U}_N(k) \quad (11)$$

where $\boldsymbol{g} = [1,0,\cdots,0]^T$.

(10) and (11) form the designed generalized predictive controller.

Performance analysis and controller design are conducted simultaneously, and (3) can be rewritten as

$$A(z^{-1})\Delta y(k+1) = B(z^{-1})\Delta u(k) + \Delta\chi(k+1) \quad (12)$$

Define

$$\boldsymbol{P} = [\tilde{\boldsymbol{\Phi}}^T \boldsymbol{Q} \tilde{\boldsymbol{\Phi}} + \boldsymbol{\lambda}]^{-1} \tilde{\boldsymbol{\Phi}}^T \boldsymbol{Q} \quad (13)$$

$$\left[ \left[\tilde{\boldsymbol{\Psi}}_1\right]_{N \times na} \quad \left[\tilde{\boldsymbol{\Psi}}_2\right]_{N \times (nb+1)} \right] = \tilde{\boldsymbol{\Psi}} \quad (14)$$

Based on (10) to (14), we can derive the following closed-loop system equation:

$$\left[ A(z^{-1})\Delta + z^{-1}B(z^{-1})(1+z^{-1}\boldsymbol{P}\tilde{\boldsymbol{\Psi}}_2 \boldsymbol{T}_u)^{-1} \boldsymbol{P}\tilde{\boldsymbol{\Psi}}_1 \boldsymbol{T}_y \Delta \right.$$
$$\left. + z^{-1}B(z^{-1})(1+z^{-1}\boldsymbol{P}\tilde{\boldsymbol{\Psi}}_2 \boldsymbol{T}_u)^{-1} \boldsymbol{P}\boldsymbol{E} \right] y(k+1) \quad (15)$$
$$= B(z^{-1})(1+z^{-1}\boldsymbol{P}\tilde{\boldsymbol{\Psi}}_2 \boldsymbol{T}_u)^{-1} \boldsymbol{P}\boldsymbol{H} y^*(k+1) + \Delta\chi(k+1)$$

where $\boldsymbol{H} = [1, z, \cdots, z^{N-1}]^T$, $\boldsymbol{T}_y = [1, \cdots, z^{-n_a+1}]^T$ and $\boldsymbol{T}_u = [1, z^{-1}, \cdots, z^{-n_b}]^T$.

We may design appropriate values for $N$, $\boldsymbol{\lambda}$, and $\boldsymbol{Q}$ to configure the system's poles to satisfy the following inequality.

$$T(z^{-1}) = \left[ A(z^{-1})\Delta + z^{-1}B(z^{-1})(1+z^{-1}\boldsymbol{P}\tilde{\boldsymbol{\Psi}}_2 \boldsymbol{T}_u)^{-1} \boldsymbol{P}\tilde{\boldsymbol{\Psi}}_1 \boldsymbol{T}_y \Delta \right.$$
$$\left. + z^{-1}B(z^{-1})(1+z^{-1}\boldsymbol{P}\tilde{\boldsymbol{\Psi}}_2 \boldsymbol{T}_u)^{-1} \boldsymbol{P}\boldsymbol{E} \right] = 0 \quad |z| > 1 \quad (16)$$

indicating that the characteristic roots are placed within the unit circle.

Assuming the control system is stable and disregarding the disturbance term $\chi(k)$, we can utilize the final value theorem to calculate the steady-state error as

$$\begin{aligned}\lim_{k\to\infty} e(k+1) &= \lim_{k\to\infty}\left[y^*(k+1) - y(k+1)\right] \\ &= \lim_{z\to 1}(1-z^{-1})T^{-1}(z^{-1})\Big[A(z^{-1})\Delta \\ &\quad + z^{-1}B(z^{-1})(1+z^{-1}P\tilde{\Psi}_2 T_u)^{-1}P\tilde{\Psi}_1 T_y \Delta \\ &\quad - B(z^{-1})(1+z^{-1}P\tilde{\Psi}_2 T_u)^{-1}P(H-z^{-1}E)\Big]Z(y^*(k+1))\end{aligned} \quad (17)$$

where $Z(\cdot)$ represents the $z$-transform.

Furthermore, we can determine the transfer function that relates the system's output to the disturbance $\chi(k)$ as

$$G_w(k) = \Delta T^{-1}(z^{-1}) \quad (18)$$

If $b_0 \neq 0$, setting $\lambda=0$, the transfer function with respect to the disturbance $\chi(k)$ is $G_w(k)=1-z^{-1}$. However, if $b_0=0$ and $\lambda$ is sufficiently small, typically $G_w(k) \neq 1-z^{-1}$.

According to (15), (17), and (18), we can use the superposition principle to predict or determine the tracking error trend of the system.

*B. Incremental Generalized Minimum Variance Controllers Design and Performance Analysis*

Suppose $b_0=\cdots=b_{d-1}=0$, indicating a delay time of $d$. When we set $N=d$, controllers (10) and (11) degenerate into incremental generalized minimum variance control. Furthermore, if we choose $Q=I$ or $Q=\text{diag}(0,\cdots,0,1)$, we will obtain the identical controller, where $\text{diag}(\cdot)$ denotes a diagonal matrix. In this case, the objective function (8) degenerates into

$$J = \left[y^*(k+d) - y(k+d)\right]^2 + \Delta U_N^T(k)\lambda\Delta U_N(k) \quad (19)$$

From (7) and (10), we obtain

$$\lambda \Delta U_N(k) = \tilde{\Phi}^T Q[Y_N^*(k+1) - Y_N(k+1) + \tilde{\Phi}_w \Delta W(k+1)] \quad (20)$$

When $\lambda=\lambda I \neq 0$, from (11), (12), and (20), we can obtain

$$\begin{aligned}\left[\lambda A(z^{-1})\Delta + B(z^{-1})g^T\tilde{\Phi}^T QH\right] y(k+1) \\ = B(z^{-1})g^T\tilde{\Phi}^T QH y^*(k+1) \\ + B(z^{-1})g^T\tilde{\Phi}^T Q\tilde{\Phi}_w H\Delta\chi(k+1) + \lambda\Delta\chi(k+1)\end{aligned} \quad (21)$$

We can choose appropriate $\lambda$ and $Q$ to configure the system's poles such that the following inequality holds:

$$T_1(z^{-1}) = \left[\lambda A(z^{-1})\Delta + B(z^{-1})g^T\tilde{\Phi}^T QH\right] \neq 0 \quad |z|>1 \quad (22)$$

, thereby ensuring that the characteristic roots remain inside the unit circle, ensuring system stability.

Assuming the control system is stable and not taking into account the influence of the disturbance term $\chi(k)$, we can apply the final value theorem to calculate the steady-state error:

$$\begin{aligned}\lim_{k\to\infty} e(k+1) &= \lim_{k\to\infty}\left[y^*(k+1) - y(k+1)\right] \\ &= \lim_{z\to 1}(1-z^{-1})T_1^{-1}(z^{-1})\lambda A(z^{-1})Z(y^*(k+1))\end{aligned} \quad (23)$$

It can be observed that the steady-state error decreases as the absolute value of the parameter $\lambda$ decreases. If $b_0 \neq 0$, theoretically, we can eliminate the steady-state error caused by the reference trajectory type by setting $\lambda=0$. For example, in cases where system stability can be ensured, considering an acceleration reference trajectory $y^*(k)=k^2$, or even higher-order reference trajectories, we can decrease the tracking error by decreasing the absolute value of $\lambda$.

On the other hand, we can derive the transfer function that relates the system output to disturbances as

$$G_w(k) = T_1^{-1}(z^{-1})[\lambda + B(z^{-1})g^T\tilde{\Phi}^T Q\tilde{\Phi}_w H]\Delta \quad (24)$$

According to (21) or (23) and (24), we can predict the trend of the tracking error or determine the tracking error of the system based on the superposition principle.

*C. Design and Performance Analysis of Generalized Predictive Controllers with Disturbance Compensation*

The choice of the objective function remains consistent with equation (8). The design of the generalized predictive controller with disturbance compensation is given by (9) and (11). However, the disturbance $\chi(k)$ and its predicted vector $\chi(k)$ in (9) denote the true values, which may not be precisely obtainable during the control process. To address this, we introduce the variable $\hat{\chi}(k)$ to represent the estimated or measured value of disturbance $\chi(k)$ and $\hat{\chi}(k)$ to represent the prediction of disturbance vector $\chi(k)$. Then (9) can be rewritten as:

$$\begin{aligned}\Delta U_N(k) = [\tilde{\Phi}^T Q\tilde{\Phi}+\lambda]^{-1}\tilde{\Phi}^T Q[Y_N^*(k+1) \\ -Ey(k)-\tilde{\Psi}\Delta x(k)-\tilde{\Phi}_w\Delta\hat{\chi}(k+1)]\end{aligned} \quad (25)$$

The generalized predictive controllers with disturbance compensation are designed by (25) and (11).

Based on (11)-(14) and (25), we can derive the closed-loop system:

$$\begin{aligned}\Big[A(z^{-1})\Delta + z^{-1}B(z^{-1})(1+z^{-1}P\tilde{\Psi}_2 T_u)^{-1}P\tilde{\Psi}_1 T_y\Delta \\ + z^{-1}B(z^{-1})(1+z^{-1}P\tilde{\Psi}_2 T_u)^{-1}PE\Big] y(k+1) \\ = B(z^{-1})(1+z^{-1}P\tilde{\Psi}_2 T_u)^{-1}PHy^*(k+1) + \Delta\chi(k+1) \\ - B(z^{-1})(1+z^{-1}P\Psi_2 T_u)^{-1}P\tilde{\Phi}_w\Delta\hat{\chi}(k+1)\end{aligned} \quad (26)$$

We can choose appropriate values of $N$, $\lambda$ and $Q$ to place the poles such that inequality (16) holds and guarantees the characteristic roots inside the unit circle.

Furthermore, when the absolute values of the elements on the diagonal of $\lambda$ are sufficiently small and system stability can be ensured, if we set $\hat{\chi}(k+1)=\chi(k+1)$, (26) can be approximated as equation (27). If $b_0 \neq 0$, we can set $\lambda=0$, and (26) will be rewritten as equation (27).

$$\begin{aligned}\Big[A(z^{-1})\Delta + z^{-1}B(z^{-1})(1+z^{-1}P\tilde{\Psi}_2 T_u)^{-1}P\tilde{\Psi}_1 T_y\Delta \\ + z^{-1}B(z^{-1})(1+z^{-1}P\tilde{\Psi}_2 T_u)^{-1}PE\Big] y(k+1) \\ = B(z^{-1})(1+z^{-1}P\tilde{\Psi}_2 T_u)^{-1}PHy^*(k+1)\end{aligned} \quad (27)$$

The transfer function of the system in relation to disturbances becomes **0**, thus theoretically eliminating the influence of disturbances on the system output.

## V. SIMPLIFICATION OF THE GENERALIZED PREDICTIVE CONTROLLER

To simplify the controller design and performance analysis, this paper has achieved the design of the aforementioned generalized predictive control, incremental generalized Minimum variance control, and GPC with disturbance compensation. However, the explicit predictive controllers mentioned above cannot be used when there is a delay ($d>1$) and when the parameter matrix $\boldsymbol{\lambda}=\mathbf{0}$, as the matrix $\tilde{\boldsymbol{\Phi}}^T\boldsymbol{Q}\tilde{\boldsymbol{\Phi}}$ becomes singular. Therefore, this section simplifies the form of generalized predictive control for such cases.

Assuming a system delay of $d$, i.e., $b_0=\cdots=b_{d-1}=0$, (28)-(30) holds true at all times.

$$\tilde{\boldsymbol{\Phi}} = diag(\underbrace{0,\cdots,0}_{d-1},1,\cdots,1)\tilde{\boldsymbol{\Phi}}diag(1,\cdots,1,\underbrace{0,\cdots,0}_{d-1}) \quad (28)$$

$$\boldsymbol{P} = diag(1,\cdots,1,\underbrace{0,\cdots,0}_{d-1})\boldsymbol{P}diag(\underbrace{0,\cdots,0}_{d-1},1,\cdots,1) \quad (29)$$

$$\Delta\boldsymbol{U}_N(k) = diag(1,\cdots,1,\underbrace{0,\cdots,0}_{d-1})\Delta\boldsymbol{U}_N(k) \quad (30)$$

For the sake of convenience, we define "×" to represent the deletion operation. Similar to matrix multiplication, we define the left multiplication by diagonal matrix $diag(\underbrace{\times,\cdots,\times}_{d-1},1,\cdots,1)$ to mean the removal of the first $d-1$ rows from the matrix, and the right multiplication by $diag(\underbrace{\times,\cdots,\times}_{d-1},1,\cdots,1)$ to mean the removal of the first $d-1$ columns from the matrix. Similarly, we define left and right multiplication operations for $diag(1,\cdots,1,\underbrace{\times,\cdots,\times}_{d-1})$.

Further, we let $\boldsymbol{\lambda}_d=diag(\lambda_1,\cdots,\lambda_{N-d+1})$, $\boldsymbol{Q}_d=diag(q_d,\cdots,q_N)$,

$\tilde{\boldsymbol{\Phi}}_d = diag(\underbrace{\times,\cdots,\times}_{d-1},1,\cdots,1)\tilde{\boldsymbol{\Phi}}diag(1,\cdots,1,\underbrace{\times,\cdots,\times}_{d-1})$,

$\tilde{\boldsymbol{\Psi}}_d = diag(\underbrace{\times,\cdots,\times}_{d-1},1,\cdots,1)\tilde{\boldsymbol{\Psi}}(k)$,

$\tilde{\boldsymbol{\Phi}}_{wd} = diag(\underbrace{\times,\cdots,\times}_{d-1},1,\cdots,1)\tilde{\boldsymbol{\Phi}}_w$,

$\boldsymbol{Y}^*_{Nd}(k+1) = diag(\underbrace{\times,\cdots,\times}_{d-1},1,\cdots,1)\tilde{\boldsymbol{Y}}^*_N(k+1)$
$= [y^*(k+d),\cdots,y^*(k+N)]^T$,

$\boldsymbol{Y}_{Nd}(k+1) = [y(k+d),\cdots,y(k+N)]^T$,

$\Delta\boldsymbol{U}_{Nd}(k) = [\Delta u(k)\cdots\Delta u(k+N-d)]$,

$\boldsymbol{E}_d = diag(\underbrace{\times,\cdots,\times}_{d-1},1,\cdots,1)\boldsymbol{E}$, $\boldsymbol{H}_d = [z^{-d+1},\cdots,z^{N-1}]^T$,

$\boldsymbol{T}_{yd} = [z^{-d+1},\cdots,z^{-n_a+1}]^T$, $\boldsymbol{T}_{ud} = [z^{-d+1},\cdots,z^{-n_b}]^T$.

Then (10) in part A can be rewritten as

$$\Delta\boldsymbol{U}_{Nd}(k) = [\tilde{\boldsymbol{\Phi}}_d^T\boldsymbol{Q}\tilde{\boldsymbol{\Phi}}_d + \boldsymbol{\lambda}_d]^{-1}\tilde{\boldsymbol{\Phi}}_d^T\boldsymbol{Q}[\boldsymbol{Y}^*_{Nd}(k+1) \\ -\boldsymbol{E}_d y(k) - \tilde{\boldsymbol{\Psi}}_d(k)\Delta\boldsymbol{x}(k)] \quad (31)$$

The incremental generalized minimum variance controller in part B can be written as

$$\Delta u(k) = \tilde{\boldsymbol{\Phi}}_d^T q_N[\boldsymbol{Y}^*_{Nd}(k+1) - \boldsymbol{E}_d y(k) \\ -\tilde{\boldsymbol{\Psi}}_d(k)\Delta\boldsymbol{x}(k)]/[\tilde{\boldsymbol{\Phi}}_d^T q_N\tilde{\boldsymbol{\Phi}}_d + \lambda_1] \\ = b_{d-1}q_N[y^*(k+d) - y(k) \\ -\tilde{\boldsymbol{\Psi}}_d(k)\Delta\boldsymbol{x}(k)]/[q_N b_{d-1}^2 + \lambda_1] \quad (32)$$

The GPC with disturbance compensation (25) can be rewritten as

$$\Delta\boldsymbol{U}_{Nd}(k) = [\tilde{\boldsymbol{\Phi}}_d^T\boldsymbol{Q}_d\tilde{\boldsymbol{\Phi}}_d + \boldsymbol{\lambda}_d]^{-1}\tilde{\boldsymbol{\Phi}}_d^T\boldsymbol{Q}[\boldsymbol{Y}^*_{Nd}(k+1) \\ -\boldsymbol{E}_d y(k) - \tilde{\boldsymbol{\Psi}}_d(k)\Delta\boldsymbol{x}(k) - \tilde{\boldsymbol{\Phi}}_{wd}\Delta\hat{\boldsymbol{\chi}}(k+1)] \quad (33)$$

Similarly, the closed-loop system equations can be obtained by replacing $\boldsymbol{\lambda}$, $\boldsymbol{Q}$, $\tilde{\boldsymbol{\Phi}}\cdots$ with the corresponding $\boldsymbol{\lambda}_d$, $\boldsymbol{Q}_d$, $\tilde{\boldsymbol{\Phi}}_d\cdots$ as mentioned above.

In addition, when we choose $N=d$, (33) can be written as (34)

$$\Delta\boldsymbol{U}_{Nd}(k) = \tilde{\boldsymbol{\Phi}}_d^T\boldsymbol{Q}[\boldsymbol{Y}^*_{Nd}(k+1) - \boldsymbol{E}_d y(k) - \tilde{\boldsymbol{\Psi}}_d(k)\Delta\boldsymbol{x}(k) \\ -\tilde{\boldsymbol{\Phi}}_{wd}\Delta\hat{\boldsymbol{\chi}}(k+1)]/[\tilde{\boldsymbol{\Phi}}_d^T\boldsymbol{Q}_d\tilde{\boldsymbol{\Phi}}_d + \boldsymbol{\lambda}_d] \quad (34)$$

which corresponds to the incremental generalized minimum variance control with disturbance compensation.

Essentially, the objective function corresponding to the controller (31)-(34) can be expressed as:

$$J = \left[\boldsymbol{Y}^*_{Nd}(k+1) - \boldsymbol{Y}_{Nd}(k+1)\right]^T \boldsymbol{Q}_d \left[\boldsymbol{Y}^*_{Nd}(k+1) - \boldsymbol{Y}_{Nd}(k+1)\right] \\ + \Delta\boldsymbol{U}^T_{Nd}(k)\boldsymbol{\lambda}_d\Delta\boldsymbol{U}_{Nd}(k) \quad (35)$$

or be written as

$$J = \left[\boldsymbol{Y}^*_N(k+1) - \boldsymbol{Y}_N(k+1)\right]^T \boldsymbol{Q}_0 \left[\boldsymbol{Y}^*_N(k+1) - \boldsymbol{Y}_N(k+1)\right] \\ + \Delta\boldsymbol{U}^T_N(k)\boldsymbol{\lambda}_0\Delta\boldsymbol{U}_N(k) \quad (36)$$

where $\boldsymbol{Q}_0 = diag(\underbrace{0,\cdots,0}_{d-1},1,\cdots,1)\boldsymbol{Q}$ and $\boldsymbol{\lambda}_0 = \boldsymbol{\lambda}diag(1,\cdots,1,\underbrace{0,\cdots,0}_{d-1})$.

Furthermore, by setting the parameter matrix $\boldsymbol{\lambda}_d=\mathbf{0}$, under the condition of system stability and no disturbance or with disturbance compensation $\hat{\boldsymbol{\chi}}(k+1) = \boldsymbol{\chi}(k+1)$, we can obtain the characteristic equation of the system.

$$diag(1,\cdots,1,\underbrace{\times,\cdots,\times}_{d-1})\boldsymbol{Y}^*_N(k+1) = diag(1,\cdots,1,\underbrace{\times,\cdots,\times}_{d-1})\boldsymbol{Y}_N(k+1) \quad (37)$$

where (37) involves the reference trajectory and predicted output at future times, while for (disturbance-compensated) incremental generalized minimum variance control, the characteristic equation of the system can be obtained as:

$$y(k+1) = y^*(k+1) \quad (38)$$

For reference trajectories $k^n$ ($n=1,2,\cdots$), theoretically achieving error-free tracking, is crucial. According to the Taylor series, for any order-differentiable function $F(k) = \sum_{i=1}^{\infty}\frac{F^{(n)}(k)}{n!}k^i$. If we want the system to be able to track the reference trajectory $y^*(k)=F(k)$ without steady-state error, the controller should, at least in theory, have the ability to make the system track $k^n$ ($n=1,2,\cdots,n_{max}$) without steady-state error. $n_{max}$ is the highest power of $k$ when $F(k)$ is expanded into a Taylor series. In educational examples, cubic polynomials, quintic polynomials, and septenary polynomials are commonly used to illustrate the generation of trajectories [36].

Remark 1: Typically, the computational complexity of matrix inversion increases with the dimensionality of the matrix. In Section V, the order of the matrix $[\tilde{\boldsymbol{\Phi}}_d^T\boldsymbol{Q}_d\tilde{\boldsymbol{\Phi}}_d + \boldsymbol{\lambda}_d]$ involved in generalized predictive control is $N-d+1$, which is smaller by d-1 compared to the matrices that need to be inverted in [9], [21] and [25]. This reduction in dimensionality leads to a significant advantage as it reduces the computational workload required for solving the controller. Furthermore, when $N=d$, the controller degenerates into the incremental generalized minimum variance control, thereby eliminating the need for matrix inversion operations.

In [25], the GPC introduces the control time domain $N_u$ to reduce computational load. This can be achieved by setting $\boldsymbol{\Psi}_{Nu}=\boldsymbol{\Psi}_{Nd}\boldsymbol{K}$, where $\boldsymbol{K}=diag(\underbrace{1,\cdots,1,\times,\cdots,\times}_{N_u})$. In the corresponding performance analysis, you can replace $\boldsymbol{\Psi}_{Nu}$ with $\boldsymbol{\Psi}_d$. In practical simulations, when $N_u<N-d+1$, control performance often deteriorates. The authors believe that predictive control is essentially optimal control, and this artificial simplification of the computational load in the GPC changes its characteristics. Therefore, $N_u=N$ is the default setting in this paper, meaning that the control time domain is not considered.

## VI. SIMULATION EXPERIMENTS

*Example 1*: Consider the following linear system
$$y(k)-0.5y(k-1)-0.8y(k-2)=2u(k-3)+u(k-4) \\ +0.5u(k-5)+\chi(k) \quad (39)$$

The disturbance term is of Type I input (ramp input) $\chi(k)=k$. The reference is chosen as a ramp input:
$$y*(k)=k \quad 1\leq k \leq 400 \quad (40)$$

The GPC (10), (11) in Section IV is employed, with controller parameters chosen as follows: $N=4$, $\boldsymbol{Q}=\boldsymbol{I}$, $\boldsymbol{\lambda}=\lambda\boldsymbol{I}$.

The actual error in the simulation is denoted as $e(k)$. Based on the superposition principle, the steady-state error, denoted as $E(k)$, is computed using (17) and (18). Table I presents the values of $e(k)$ and $E(k)$ from time 70 to time 400 for different controller parameter values of $\lambda$. The precision is maintained up to ten decimal places.

Table 1 The values of $e(k)$ and $E(k)$

| $\lambda$ | −0.1 | 1 | 2 |
|---|---|---|---|
| $e(70)=\cdots=e(400)$ | −4.8827413127 | −5.9632653061 | −6.3657142857 |
| $E(70)=\cdots=E(400)$ | −4.8827413127 | −5.9632653061 | −6.3657142857 |

Table 1 indicates that for controller parameter values of $\lambda$ equal to -0.1, 1, and 2, the actual error values $e(k)$ and the corresponding computed steady-state error values $E(k)$ are equal from time step $k=70$ to $k=400$. This validates the conclusions from the performance analysis in Part IV. It also shows that $\lambda$ can indeed be a negative value. From the perspective of using system performance analysis (16), (17) to guide controller design, its principles can be understood. Often, we can find a trade-off between system convergence and stability by adjusting $\lambda$. Decreasing the absolute value of $\lambda$ can sometimes improve the convergence speed and error convergence accuracy of the system. However, for non-minimum phase systems, excessively small absolute values of $\lambda$ may not guarantee system stability.

*Example 2*: Consider the following linear system
$$y(k)-0.5y(k-1)-0.8y(k-2)=u(k-4) \\ +0.5u(k-5)+\chi(k) \quad (41)$$

where the disturbance term is a random noise with a mean of 0 and a variance of 1, represented as $\chi(k)=rand$ and "rand" is a MATLAB function. The reference is chosen as:

$$y*(k)=(-1)^{round(k/100)} \quad 1\leq k \leq 400 \quad (42)$$

Let $N=d=4$, i.e., the incremental generalized minimum variance controller in Section V. The controller parameters are $\boldsymbol{Q}=\boldsymbol{I}$ or $\boldsymbol{Q}=diag(0,0,0,1)$, and $\boldsymbol{\lambda}=10^{-10}\boldsymbol{I}$. The system output is shown in Fig. 1. $E(k)$ represents the tracking error caused by disturbances calculated according to (24), and Fig. 2 displays $E(k)$ and the actual tracking error $e(k)$ in the simulation.

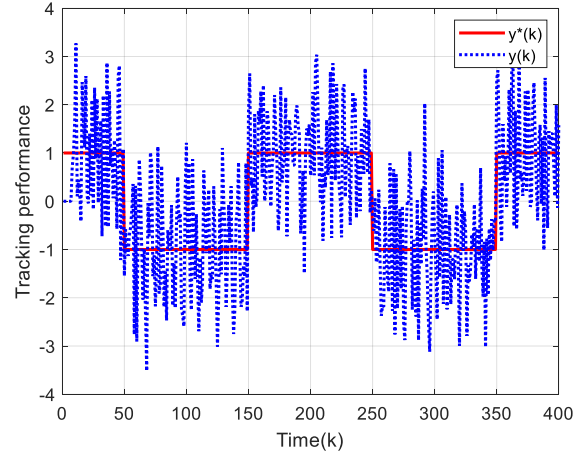

Fig. 1 Tracking performance

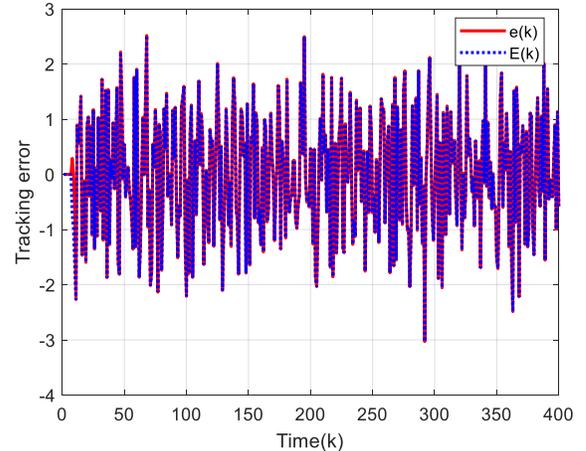

Fig. 2 Tracking error and calculated tracking error

Fig. 2 shows that the calculated tracking error $E(k)$ caused by disturbances coincides with the actual tracking error values $e(k)$, validating the correctness of the performance analysis in Section IV. In this example, due to the influence of system delays, when using generalized predictive control or incremental generalized minimum variance control, the tracking error caused by the system noise term may be larger than the noise term itself, demonstrating an amplification effect of the system noise term.

On the other hand, it is essential to investigate the tracking of control algorithms for high-order reference trajectories. [36] used the example of a joint trajectory for a robot using a fifth-degree polynomial. In practical applications, more stringent requirements may necessitate the use of higher-degree polynomials as path segments (reference trajectories). Therefore, it is important to study the tracking performance of control algorithms for high-degree reference trajectories. To verify the conclusions regarding the steady-state error of the incremental generalized minimum variance controller (32) in Section V, we set the disturbance term $\chi(k)$ to zero and select the reference trajectory as:

$$y^*(k) = k \qquad 1 \leq k \leq 400 \qquad (43)$$

When we employ the incremental generalized minimum variance controller (32) in Section V, Fig. 3 provides the system output $y(k)$ for different controller parameters $\lambda_1=0$, $\lambda_2=-1$ and $\lambda_3=5$, along with the actual tracking error $e(k)$ for $\lambda_1=0$. Fig. 4 displays the corresponding control inputs.

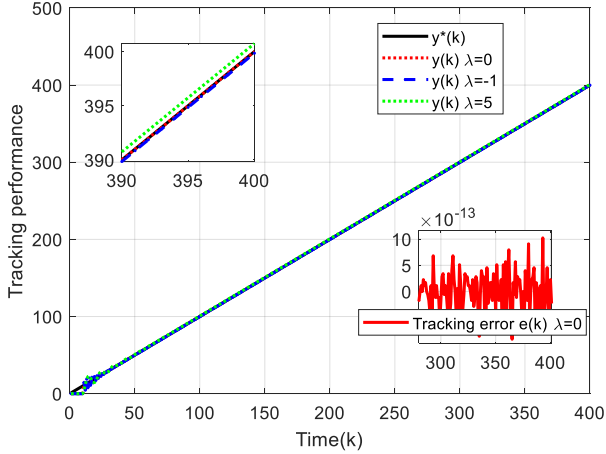

Fig. 3 Tracking performance

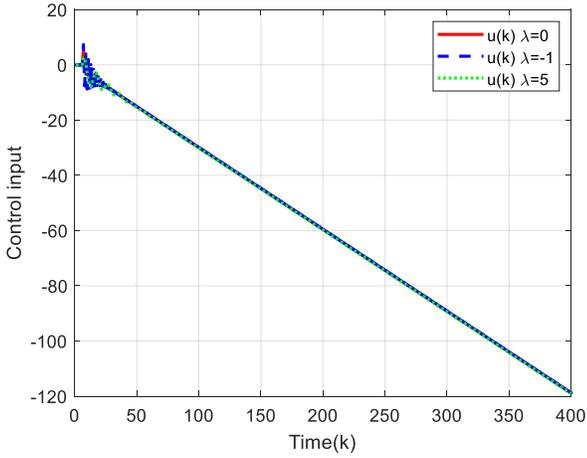

Fig. 4 Control input

From Fig. 3, it is evident that the tracking error decreases as the absolute value of $\lambda$ decreases. When the controller parameter $\lambda_1=0$, the tracking error from $k=300$ to $k=400$ is in the range of $-5\times10^{-10}<e(k)<5\times10^{-10}$, which is much smaller than the magnitude of the reference trajectory and can be considered negligible. Therefore, in cases where system stability is guaranteed, we can reduce the absolute value of $\lambda$ to decrease tracking errors and expedite convergence. On the other hand, by adjusting the sign of the controller parameter $\lambda$, we can alter the sign of the steady-state error, thereby determining whether the system output tracks the reference trajectory from above or below.

*Example 3*: Consider the system model (39), where the disturbance term is $\chi(k)=k^3$. We select the reference trajectory (42) and use the generalized predictive controller with disturbance compensation (25) and (11). Let $\hat{\chi}(k+1) = \chi(k+1)$, $N=4$, $Q=I$ and $\lambda=10^{-10}I$. The system output is shown in Fig. 5, and the corresponding control input is depicted in Fig. 6.

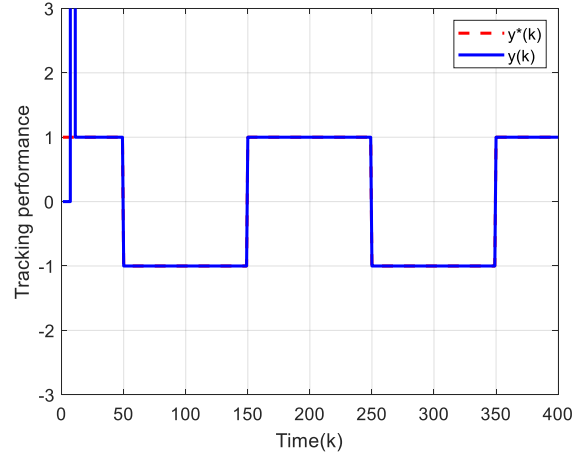

Fig. 5 Tracking performance

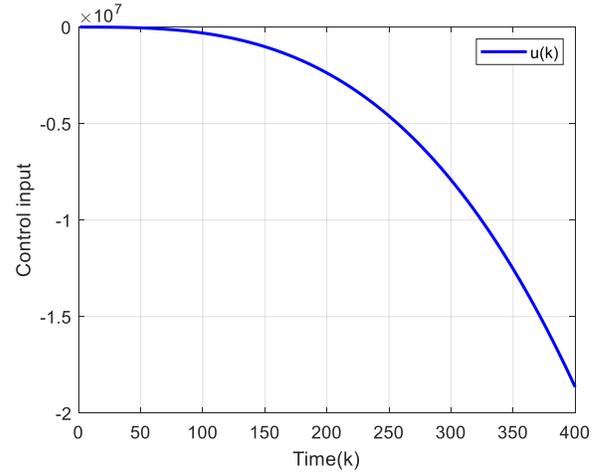

Fig. 6 Control input

The tracking performance shown in Fig. 5 demonstrates that when the disturbance term is known and predictable, and the system stability is guaranteed, using the generalized predictive controller with disturbance compensation, along with a sufficiently small absolute value of $\lambda$, can effectively reduce the impact of disturbances on the output. Additionally, the predictive controller minimizes the tracking error of the system.

## VII. CONCLUSIONS

This paper reintroduces a class of generalized predictive control and analyzes the system characteristic equation and steady-state behavior. Finally, the analysis results are validated through simulations.